\documentclass[aps,prl,twocolumn,10pt,showpacs,superscriptaddress,amsmath,amssymb,nobibnotes]{revtex4-1}

\usepackage{graphicx}
\usepackage{amsfonts}    
\usepackage{graphicx}   
\usepackage{verbatim}   
\usepackage{color}      
\usepackage{subfigure}  
\usepackage{hyperref}   
\usepackage{natbib}
\usepackage{booktabs}
\usepackage{threeparttable}
\usepackage{dcolumn}
\usepackage{multirow}
\usepackage{float}
\usepackage{ulem}

\begin{document}


\title{Quantum Enhanced Interferometer for Kilohertz Gravitational Wave Detection }

\author{Meng-Jun Hu}
\email{humj@baqis.ac.cn}
\affiliation{Beijing Academy of Quantum Information Sciences, Beijing 100193, China}

\author{Shuai Zha}
\affiliation{The Oskar Klein Centre, Department of Astronomy, Stockholm University, AlbaNova, SE-106 91 Stockholm, Sweden}

\author{Yong-Sheng Zhang}%
\email{yshzhang@ustc.edu.cn}
\affiliation{Laboratory of Quantum Information, University of Science and Technology of China, Hefei 230026, China }%
\affiliation{Synergetic Innovation Center of Quantum Information and Quantum Physics,
University of Science and Technology of China, Hefei 230026, China}

\date{\today}

\begin{abstract}
The gravitational wave detector of higher sensitivity and greater bandwidth in kilohertz window is required for future gravitational wave astronomy and cosmology. Here we present a new type broadband high frequency laser interferometer gravitational wave detector utilizing polarization of light as signal carrier. Except for Fabry-Perot cavity arms we introduce dual power recycling to further amplify the gravitational wave signals. A novel method of weak measurement amplification is used to amplify signals for detection and to guarantee the long-term run of detector. Equipped with squeezed light, the proposed detector is shown sensitive enough within the window from 300Hz to several kHz, making it suitable for the study of high frequency gravitational wave sources. 
With the proposed detector added in the current detection network, we show that the ability of exploring binary neutron stars merger physics be significantly improved.
The detector presented here is expected to provide an alternative way of exploring the possible ground-based gravitational wave detector for the need of future research.
\end{abstract}

\maketitle


\section{I: Introduction}

The direct detection of gravitational waves from two binary black holes by Advanced Laser Interferometer Gravitational-wave Observatory (LIGO) in 2016 opens new era for astronomy and cosmology \cite{Ligo}. At the end of the second observing run (O2), a total of 11 confident gravitational waves events have been confirmed by LIGO and Virgo collaborations \cite{O2}. Among these events, the detection of binary neutron star inspiral GW170817 \cite{BNS} with a follow-up electromagnetic identification \cite{E1, E2, E3, E4, E5} significantly advances the development of multi-messenger astronomy \cite{2017multi}. 
In order to fulfill the requirements of research in next decades, the ground-based gravitational waves detection need to follow two development paths simultaneously. One path is to build a global detection network such that more detections with greater accuracy are obtained. A preliminary detection network composed of LIGO, Virgo, GEO600 and KAGRA has been formed \cite{network}, in which more detectors will join in the future. 
The another path is to improve detection sensitivity by upgrading current detectors or building more advanced detectors \cite{build}. 


In the recent third observing run (O3), the squeezed light was injected into the dark port of interferometers by both LIGO and Virgo, improving the sensitivity of detectors to signals by up to 3dB in shot-noise limited frequency range, which leads to an obvious boost for detection range and rate\cite{S1, S2}. Further application of frequency-dependent squeezing for brand-band reduction of quantum noise is also possible \cite{fdsqueezing1, fdsqueezing2, fdsqueezing3}. 
Besides the direct impact on the ability to detect astrophysical sources, the enhanced sensitivity in high frequency is also of importance for research of detected sources. 
For example, sky location is essential to multi-messenger astronomy and its accuracy highly depends on high frequency sensitivity of detectors \cite{sky}. 
Information of states of neutron stars e.g., tidal deformability and interior structure, is also carried by high frequency gravitational wave signals \cite{state}.  
In addition, gravitational waves emitted by sources relating to them such as gravitational collapse, rotational instabilities and oscillations of the remnant compact objects are expected to be within high frequency window \cite{source1, source2, source3, source4, source5, source6, source7, source8, source9, source10, source11, source12, source13, source14}.
There is thus a strong motivation for the development of ground-based gravitational wave detectors with high frequency detection window \cite{detector1, detector2, detector3, detector4,detector5, detector6}. Kip Throne ever said that ``{\it As experimental gravity pushes toward higher and higher precision, it has greater and greater need of new ideas and technology from the quantum theory of measurement, quantum optics, and other branches of physics, applied physics, and engineering}" \cite{Kip}. 
New detection methods and technologies thus have to be proposed and used for the next generation gravitational wave detection.

\begin{figure*}[tbp]
\centering
\includegraphics[scale=0.55]{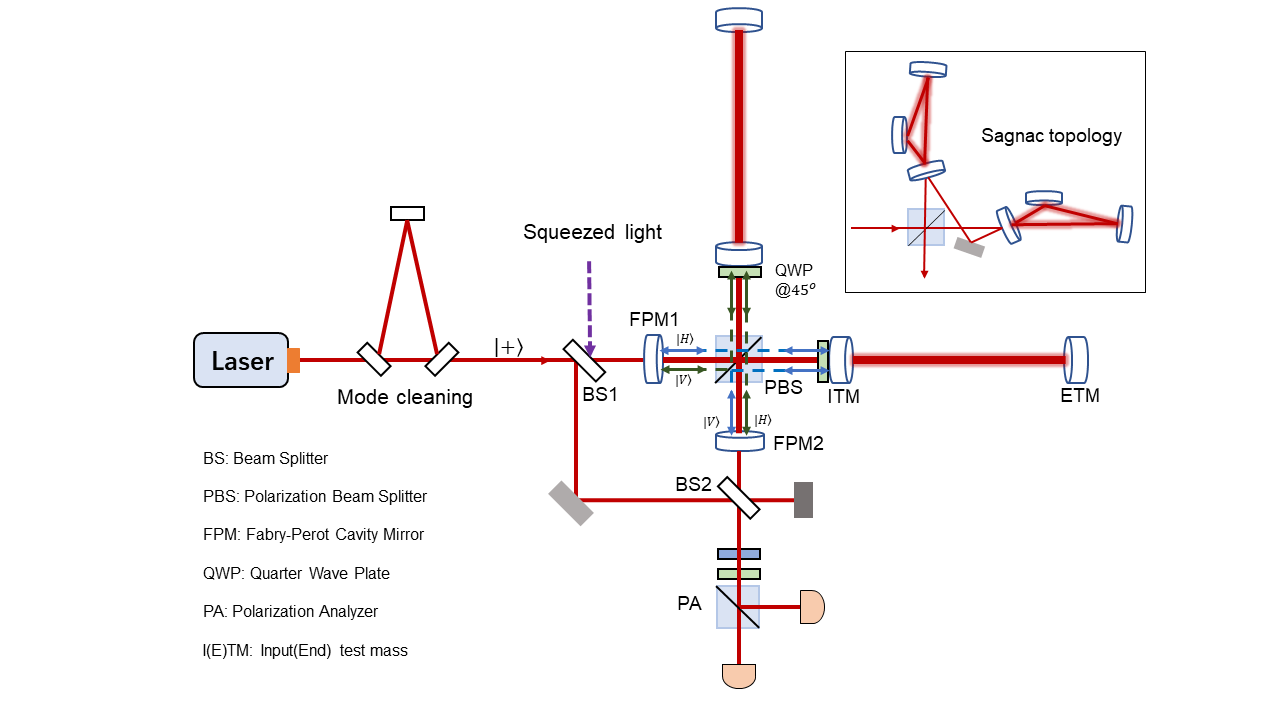}
\caption{ {\bf Schematic diagram of proposed brandband high frequency laser interferometer gravitational wave detector.} The basic configuration of the detector can be viewed as a polarization Michelson interferometer nested in an asymmetry Mach-Zehnder interferometer. BS1 is a 50:50 beam splitter, while BS2 is slight deviation from 50:50 to realize weak measurement amplification. Except Fabry-Perot cavity arms, FPM1 and FPM2 with the same transmissivity and reflectivity are used as dual power recycling cavity to amplify gravitational-wave signals further. The input light is prepared in the polarization state $|+\rangle=(|H\rangle+|V\rangle)/\sqrt{2}$ with $|H\rangle, |V\rangle$ represent horizontal and vertical polarization state respectively. Gravitational wave signals are encoded in the polarization state of light and can be extracted via a polarization analyzer placed in the dark port of BS2. Squeezed light can be injected into the dark port of BS1 and the Michelson topology can be replaced by zero-area Sagnac topology to realize quantum non-demolition measurement. }
\label{f1}
\end{figure*} 

In this paper, we propose a new type broadband laser interferometer gravitational wave detector with high frequency window, in which polarization of light is utilized as signal carrier. 
The antenna of detector is a polarization Michelson interferometer with Fabry-Perot cavity arms. 
In order to significantly reduce the shot noise in high frequency range, dual power recycling is introduced. 
A novel method called weak measurement amplification \cite{Hu} is adopted to amplify signal for detection and to guarantee the long-term run of the detector.   
The calculation of quantum noise spectrum shows this detector with current technology is sensitive enough to gravitational waves within the range from above 300Hz to several kHz, making it suitable for the study of high frequency gravitational wave astronomy and cosmology. 
The possibility of using squeezed light and quantum non-demolition measurement to surpass the quantum standard limit in proposed detector are also discussed. 

\section{II: Description of the detector}
The schematic diagram of proposed laser interferometer gravitational wave detector is shown in Fig. \ref{f1}. 
The basic optical configuration of detector can be considered as a polarization Michelson interferometer nested in an asymmetry Mach-Zehnder interferometer in which the second beam splitter BS2 is not strictly 50:50 but has slight deviation to realize weak measurement amplification (see Appendix A for details).   
The polarization Michelson interferometer with Fabry-Perot cavity arms is placed in the middle part of another Fabry-Perot cavity consists of mirrors FPM1 and FPM2. The FPM1 and FPM2 have the same reflection and transmission coefficients such that the input light from BS1 will totally output from FPM2 under resonance. Since the cavity supports two modes of polarization, we call it dual power recycling, which is different from the standard dual recycling Fabry-Perot Michelson interferometer  configuration used in LIGO/Virgo.
In order to extract signals encoded in the polarization state, light that come out from the dark port of BS2 is measured via polarization analyzer in the circular basis. 
Most of light in the LIGO detector is reflected back toward laser source, while in our case it comes out of bright port of BS2 and thus the parasitic interference in the input chain is eliminated. 
The highly resemblance of optical configuration to LIGO/Virgo detectors implies that many advanced technologies developed in LIGO/Virgo can be directly used in proposed detector.

When gravitational waves passing through the detector, a tiny changes of relative length is introduced and the signal will be first amplified by Fabry-Perot cavities in two interferometer arms \cite{michelson}. The cavity can be considered as a reflection mirror with equivalent reflection coefficient of
\begin{equation}
    r_{FP}(L) = \frac{r_{ITM}+e^{-2ik\cdot L}}{1+r_{ITM}e^{-2ik\cdot L}},
\end{equation}
where $r_{ITM}$ represents the reflection coefficient of input test mass and $L$ is the cavity length. Under the resonance condition of $e^{-2ik\cdot L_{res}}=-1$, we have $r_{FP}(L_{res})=-1$ that describes a total reflection. A small deviation from the resonance length $L_{res}$ causes
\begin{equation}
    r_{FP}(L_{res}+\delta L) \approx -e^{-2ik\cdot G_{arm}\cdot\delta L},
\end{equation}
where the higher order terms is omitted because $\delta L$ caused by gravitational wave is usually below the order of $10^{-18}$, and $G_{arm}= (1+r_{ITM})/(1-r_{ITM})$ is the gain of arm cavity. The phase of reflection light caused by $\delta L$ under resonance thus can be significantly amplified by choosing $r_{ITM}$ properly. 

We now consider the input-output relation of dual power recycling cavity consist of FPM1 and FPM2. The input light is produced in linear polarization $|+\rangle = (|H\rangle +|V\rangle)/\sqrt{2}$ with $|H\rangle$ and $|V\rangle$ represent horizontal and vertical polarization state respectively. In the cavity, photons with $|H\rangle$ or $|V\rangle$ state experience different paths as shown in Fig.\ref{f1}. We consider the input-output relation of photons with independent polarization state first and the final state of output photons can be obtained according to superposition law of optical field. Under the stationary condition, the light field within and output cavity are determined by
\begin{equation}
\begin{split}
E^{M}_{H/V} &= t\cdot E^{in}_{H/V}-r^{2}r^{2}_{FP}(L_{e/n})\cdot E^{M}_{H/V}e^{ik\cdot 2(l_{1}+l_{2})}; 
\\
E^{out}_{H/V} &= t\cdot r_{FP}(L_{e/n})\cdot E^{M}_{H/V}e^{ik(l_{1}+l_{2})},
\end{split}
\end{equation}
where $t, r$ are coefficients of transitivity and reflectivity of FPM1/FPM2, $L_{e/n}$ represent length of east or north arms in Fig.\ref{f1}, $l_{1}$ labels the distance between FPM1 and ITM and $l_{2}$ labels the distance betweem ITM and FPM2.
Choosing proper $l_{1}, l_{2}$ such that $e^{ik\cdot 2(l_{1}+l_{2})}=-1$, we obtain the equivalent transmission coefficient of FPM1-FPM2 cavity as
\begin{equation}
t_{H/V} = \frac{it^{2}\cdot r_{FP}(L_{e/n})}{1-r^{2}r^{2}_{FP}(L_{e/n})}.
\end{equation}
A tiny deviation from resonance $L_{e/n}=L_{res}$ results in
\begin{equation}
\begin{split}
t_{H/V}(L_{res}+\delta L_{e/n}) \approx -ie^{-ik\cdot G_{P}\cdot G_{arm}\cdot 2\delta L_{e/n}},
\end{split}
\end{equation}
where $G_{dp}\equiv (1+r^{2})/t^{2}$ gives the gain of dual power recycling. 
The signal of gravitational waves $h(t)$ propagating perpendicular to detector introduces relative length change of $\Delta L=\delta L_{e}-\delta L_{n} = h(t)L$. 
According to superposition rule, the polarization state of output light reads
\begin{equation}
|\varphi\rangle = \frac{1}{\sqrt{2}}(|H\rangle + e^{i\theta(t)}|V\rangle),
\end{equation}
where $\theta(t)=G_{dp}\cdot G_{arm}\cdot k\cdot 2h(t)L$ and global phase is omitted.
The detector equipped with Fabry-Perot cavity arms and dual power recycling thus amplify gravitational wave signals by the factor of $G_{dp}\cdot G_{arm}$. 
It should be noted that power recycling used in LIGO/Virgo detectors enhances only input light power rather than signal itself.

In order to catch the possible gravitational waves events, detectors need to run from several months to years in practical case. It is thus impossible to extract signals directly at the output port of FPM2. The difficulty, however, can be overcome by using weak measurement amplification (see Appendix for details). 
In proposed detector as shown in Fig. \ref{f1}, it is realized by a simple Mach-Zehnder interferometer consisting of BS1 and BS2. 
After passing through BS1, the state of photons is described as
\begin{equation}
|\Psi_{i}\rangle = (r_{1}|d\rangle+t_{1}|u\rangle)\otimes |+\rangle,
\end{equation}
where $r_{1}, t_{1}$ are coefficients of reflection and transmission of BS1 satisfying $|r_{1}|^{2}+|t_{1}|^{2}=1$ and $|d\rangle, |u\rangle$ represent path state of down and up arm respectively. The gravitational wave signals are encoded in the polarization state of photons flying along up arm. Since global phase can be compensated in the down arm, the state of photons, before arriving at BS2, reads
\begin{equation}
|\Psi_{f}\rangle = r_{1}|d\rangle\otimes |+\rangle +t_{1}|u\rangle\otimes |\psi\rangle,
\end{equation}
where $|\psi\rangle$ is shown in Eq. (6). When only photons out the down port of BS2 are considered, path state $|\phi\rangle=r_{2}|d\rangle+t_{2}|u\rangle$ is post-selected, where $r_{2}, t_{2}$ are coefficients of reflection and transmission of BS2. The polarization state of post-selected photons, in the first order approximation, becomes
\begin{equation}
|\psi\rangle=\frac{1}{\sqrt{P}}\langle\phi|\Psi\rangle_{f}=\frac{1}{\sqrt{2}}(|H\rangle+e^{i\Theta(t)}|V\rangle).
\end{equation}
Here $P=|\langle\phi|\Psi\rangle_{f}|^{2}=(r_{1}r_{2}+t_{1}t_{2})^{2}$ is the successful probability of post-selection and $\Theta(t)$ is determined by the formula of
\begin{equation}
\mathrm{tan}\Theta=\frac{\mathrm{sin\theta}}{\mathrm{cos}\theta+r_{1}r_{2}/t_{1}t_{2}}=\frac{\theta}{1+r_{1}r_{2}/t_{1}t_{2}}\equiv A\theta,
\end{equation}
where $\theta\ll 1$. Choosing $r_{1}, t_{1}, r_{2}, t_{2}$ properly such that $r_{1}r_{2}+t_{1}t_{2}\to 0$, $A > 1$ is obtained and signal $\theta$ is amplified. The amplification of signal is at the cost of low probability of success $P=(t_{1}^{2}t_{2}^{2})/A^{2}$. In proposed detector, BS1 is set that $t_{1}^{2}=r_{1}^{2}=1/2$. Choosing $r_{2}^2=\mathrm{sin}^{2}(\pi/4+\delta), t_{2}^{2}=\mathrm{cos}^{2}(\pi/4+\delta)$ for BS2 and notice that $\pi/2$ phase is added to reflection field we have $A=1/\delta$. The degree of deviation of BS2 from 50:50 beam splitter determines the ability of amplification. 

Weak measurement amplification is of importance in proposed detector for guaranteeing dark port detection such that long-term run of detector is possible. With current technology, $A=1/\delta\approx 10^{3}$ is available and the measured light intensity will be reduced to about millionth of input light from laser source. In addition, amplified signals is more robust to the calibration error of polarization analyzer. A proof of principle experimental demonstration of weak measurement amplification has already been given \cite{zhang}.
Amplified signals $\Theta(t)$ is easily extracted in the circular basis of polarization $\lbrace|R/L\rangle=(|H\rangle \pm i|V\rangle)/\sqrt{2}\rbrace$ as
\begin{equation}
\langle\psi|\hat{\sigma}|\psi\rangle=\frac{I_{1}-I_{2}}{I_{1}+I_{2}}=\mathrm{sin}\Theta,
\end{equation}
where $\hat{\sigma}\equiv|R\rangle\langle R|-|L\rangle\langle L|$ and $I_{1}, I_{2}$ are light intensities of two detection ports. The measured quantity is directly proportional to signals, while in Advanced LIGO the DC readout is required \cite{DC}.   

\section{III: Quantum Noise Analysis}
The noise spectrum determines ultimately the sensitivity and bandwidth of gravitational-wave detectors. Except quantum noise, there are various of other noises source e.g., thermal noise, seismic noise, coating noise etc.  With current state-of-art technologies, quantum noise has been the dominant noise in LIGO detectors and thus we only focus on it here.  

\begin{figure}[tbp]
\centering
\includegraphics[scale=0.28]{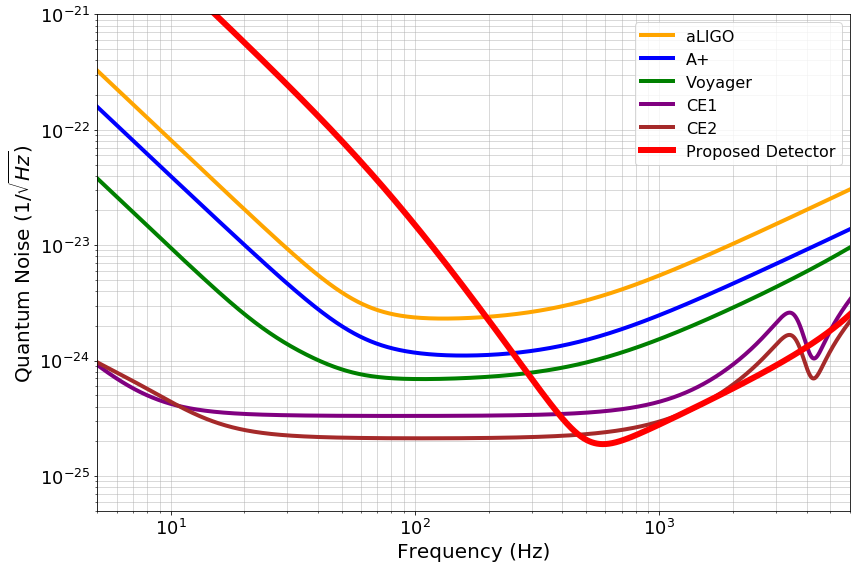}
\caption{ {\bf Quantum noise limited strain sensitivity of proposed detector vs LIGO detectors.} Parameters of proposed detector are chosen as $\lambda_{L}=1064\mathrm{nm}, P_{0}=125\mathrm{W}, L=4\mathrm{km}, M=40\mathrm{Kg}, G_{arm}=150, G_{dp}=50$. A 10dB squeezing is used here. }
\label{f2}
\end{figure} 

Quantum noise consists of radiation pressure noise and shot noise, they are dominant in different frequency domain. Radiation pressure noise takes responsibility for low frequency performance of detector, while shot noise determines the high frequency performance.  
The calculation of quantum noise in proposed detector is very similar to that in LIGO detector \cite{Miggiore, quantumnoise, KLM} except that polarization of photons should be taken into consideration. 
Notice that the minimal detectable amplified phase $\Theta$ is proportional to $1/\sqrt{2PN_{0}}$ with $P$ is the successful probability of post-selection and $N_{0}$ represents photon numbers of input light during observation time. Since the response function of Fabry-Perot cavity arm is the same in LIGO detector \cite{Miggiore}, the strain sensitivity due to quantum noise in our detector reads
\begin{equation}
\begin{split}
&S_{n}^{1/2}(f_{gw})|_{shot}=\sqrt{\frac{2\pi\hbar\lambda_{L}c}{t_{1}^{2}t_{2}^{2}P_{0}}}\cdot\frac{1}{4\pi G_{dp}\cdot G_{arm}\cdot L}\cdot\frac{1}{C(f_{gw})}      \\
&S_{n}^{1/2}(f_{gw})|_{rad} =\sqrt{\frac{\hbar t_{1}^{2}P_{0}}{2\pi\lambda_{L}c}}\cdot\frac{8\sqrt{2}\pi G_{dp}\cdot G_{arm}}{(2\pi f_{gw})^{2}\cdot M\cdot L}\cdot C(f_{gw}),
\end{split}
\end{equation}
where $C(f_{gw})$ is the frequency response function
\begin{equation}
C(f_{gw}) = \frac{\mathrm{sinc}(2\pi f_{gw}L/c)}{\sqrt{1+G_{arm}^{2}/2\cdot[1-\mathrm{cos}(4\pi f_{gw}L/c)]}}
\end{equation}
and the total strain sensitivity is $S^{1/2}_{n}(f_{gw})=\sqrt{S_{n}(f_{gw})|_{shot}+S_{n}(f_{gw})|_{rad}}$.  
When $f_{gw}\ll c/2\pi L\approx 12\mathrm{kHz} $, $C(f_{gw})$ reduces to the familiar form of $1/\sqrt{1+(f_{gw}/f_{p})^{2}}$ with $f_{p}$ is the pole frequency. Weak measurement amplification does not improve strain sensitivity of detector. The intriguing thing is that it provides us a way to extract signals without need to detect all the photons but in the meantime it maintains the sensitivity. 

The strain sensitivity of proposed detector due to quantum noise is shown in Fig. \ref{f2} with the LIGO detectors as comparison. The improvement is due to the use of dual power recycling with gain $G_{dp}$ as clearly shown in Eq. (12). In fact, if $G_{dp}$ is replaced by $\sqrt{2G_{p}}$ in Eq. (12) with $G_{p}$ is the gain of power recycling, we obtain the strain sensitivity of LIGO detector without signal recycling. With current technology it is a reliable assumption that the magnitude of $G_{dp}$ can be comparable to $G_{p}$. The squeezed light, which has already been used in LIGO/Virgo detectors, can also be equipped with proposed detector. Contrary to LIGO detector, the origin of quantum noise in proposed detector results from vacuum fluctuations of the dark port of BS1 (see Appendix B for details). The location separation of squeezed light injection from signal detection as shown in Fig. \ref{f1} is obviously beneficial for better squeezing quality and control. With current state-of-art squeezing technology \cite{squeeze1, squeeze2, squeeze3}, 10 dB squeezing in proposed detector seems plausible in practice.
The application of frequency-dependent squeezing is also possible to reduce quantum noise in low frequency.

Since gravitational wave signals are extremely weak, the behaviour of test mass is actually governed by quantum theory in which Heisenberg uncertainty principle sets a minimum noise called standard quantum limit(SQL) \cite{SQL} $S^{1/2}_{SQL}=\sqrt{8\hbar/M}/(2\pi f_{gw}L)\approx 1.83\times 10^{-22}/f_{gw}$.
The application of squeezed light discussed above is one way to beat the SQL. Another way is the quantum non-demolition measurement (QNM) \cite{QNM}, with which detectors will only be limited by shot noise.  Laser interferomter with zero-area Sagnac topology configuration as a speed meter is considered one of promising ways to realize QNM of gravitational wave signals \cite{chen1, chen2, chen3}.  
The proposed detector can be naturally redesigned with zero-area Sagnac topology as shown in block diagram of Fig. \ref{f1}. 
It is believed that technical issues will not be limiting factor and the QNM detectors shall be next generation ground-based gravitational-wave detectors \cite{n1}. 

\section{IV: Astronomy Applications}
The proposed detector with significantly improved sensitivity in high frequency will contribute a lot to the exploration of high frequency gravitational wave source e.g., merger physics of binary neutron stars (BNS) \cite{BNS1, BNS2}.
In order to evaluate the ability of proposed detector in exploring the physics of BNS merger, we consider such 4km long detector located in Hefei of China and join the current detection network consists of two LIGO detectors and Virgo detector. We consider a GW170817 event like gravitational wave signal emitted by a $1.5M_{\bigodot}-1.3M_{\bigodot}$ BNS located at a distance of $100\mathrm{Mpc}$ and compare the ability of detection network with and without proposed detector using open source software Bilby \cite{bilby}. The waveform IMRPhenomD$\_$NRTidal is used in calculations starting at $40\mathrm{Hz}$.
In the late-inspiral phase of BNS, tidal effects become important at frequencies $\geq 500\mathrm{Hz}$ and the information of equation of state can be extracted from tidal deformability quantified by $\widetilde{\Lambda}$ and $\delta\widetilde{\Lambda}$ \cite{tidal}. The fractional error in the tidal deformability $\Delta\widetilde{\Lambda}/\widetilde{\Lambda}$ determined by detection network constrain the possible equation of state \cite{BNS1}. 
Following the merger of BNS, a massive neutron star may survives and gravitational waves will be emitted at frequencies $1\mathrm{k}-4\mathrm{kHz}$ lasting for hundreds of milliseconds \cite{merger}. 
The reconstruction of post-merger gravitational waves provides us opportunity to explore the structure of neutron star and nuclear equation of state. With the proposed detector join in the current detection network, the combined signal-to-noise ratio (SNR) $\rho$ is significantly improved as shown in TABLE \ref{my_label} and signal is thus more easily reconstructed from background noise (see Appendix C for more details).  

\begin{table}[t]
\renewcommand\arraystretch{1.6}
\setlength\tabcolsep{6pt}
\centering
\scalebox{1.0}{
\begin{tabular}{|c|c|c|c|}
\hline
Network& A & B & C  \\
\hline
Combined signal-to-noise ratio $\rho$ &26.67 &32.24 &75.12 \\
\hline
Tidal deformability error $\Delta\widetilde{\Lambda}/\widetilde{\Lambda}$&0.038&0.031&0.005   \\
\hline
\end{tabular}}
\caption{Comparison of three different detection networks for a $1.5M_{\bigodot}-1.3M_{\bigodot}$ binary neutron stars located at distance of $100\mathrm{Mpc}$ with $\Lambda_{1}=400$ and $\Lambda_{2}=450$. A: Current network with two LIGO detectors and Virgo detector; B: Current network plus LIGO-like detector located in Hefei; C: Current network plus proposed detector located in Hefei.}
\label{my_label}
\end{table}

\section{V: Discussion and Conclusion}
Compared with current LIGO/Virgo detectors, the proposed detector has three advantages of technical implementation mainly due to the application of weak measurement amplification. The first one is the spatial separation of squeezing injection and detection port. The second one is that most of light coming out of detection system rather reflected back toward to the light source and the third one is the extraction of signals encoded in the polarization of photons without need of DC readout. 
On the other hand, however, it requires extreme high quality of manipulating polarization optics, which is a technical challenges. Detailed theoretical analysis and prototype experiment are thus needed for performance assessment in the next step.

In conclusion, a new type broadband laser interferometer gravitational-wave detector is proposed. Except for Fabry-Perot cavity arms, dual power recycling is introduced for the amplification of signals that encoded in the polarization of photons. A novel method weak measurement amplification is used to amplify signals for dark port detection such that long-term run of detector is guaranteed. Squeezed light can be injected in the detector for further reduction of quantum noise. With zero-area Sagnac topology, the detector has the potential to realize quantum non-demolition measurement.
The proposed detector is shown sensitivity enough in the high frequency within the range from above 300Hz to several kHz, making it suitable for the study of high frequency sources such as gravitational collapse, merger of binary neutron stars, rotational instabilities and oscillations of the remnant compact objects. 
With this kind of detector join the current global detection network, not only the detection range and rate can be increased but also the accuracy of sky location, which is essential to multi-messenger astronomy, can be significantly improved. 
Application of proposed detector in fundamental physics such as probing the Planck scale \cite{holometer1, holometer2} and axion dark matter \cite{axion} seems also promising.

\section{acknowledgments}
Meng-Jun Hu and Yong-Sheng Zhang acknowledge Jinming Cui, Tan Liu for helpful discussions. Meng-Jun Hu is supported by the Beijing Academy of Quantum Information Sciences.
Yong-Sheng Zhang is supported by the National Natural Science Foundation of China (No. 11674306 and No. 61590932), the Strategic Priority Research Program (B) of the Chinese Academy of Sciences (No. XDB01030200) and National key R \& D program (No. 2016YFA0301300 and No. 2016YFA0301700). In memory of those brave people who protect the world during the outbreak of COVID-19. 

\appendix
\section{Appendix A: Weak Measurement Amplification}
In this section, we will give a detailed introduction of weak measurement amplification that is essential to the proposed gravitational wave detector. The concept of weak measurement is first proposed by Aharonov, Albert and Vaidman (AAV) \cite{WV1}, which focuses on disturbing a system as small as possible so that the state of system would not collapse after measurement.

Consider a two-level system initially prepared in the superposition state of  $|\psi_{i}\rangle_{s} = \alpha|0\rangle_{s} + \beta|1\rangle_{s}$ with $|\alpha|^{2}+|\beta|^{2}=1$. For simplicity, we choose a qubit state as pointer with initial state $|0\rangle_{p}$. The initial state of the composite system is thus $|\Psi_{i}\rangle_{sp}=|\psi_{i}\rangle_{s}\otimes|0\rangle_{p}$. The interaction Hamiltonian usually takes the form of von Neumann-type as $\hat{H}=g\hat{A}\otimes\hat{\sigma}_{y}$, where $\hat{A}\equiv|0\rangle\langle 0|-|1\rangle\langle 1|$ is the observable of system, $\hat{\sigma}_{y}$ is Pauli operator acting on pointer and $g$ represents the coupling between system and pointer. According to quantum measurement theory, the system and pointer are entangled after interaction and the state of the composite system becomes
\begin{equation}
\begin{split}
&|\Psi_{f}\rangle_{sp}=e^{-ig\Delta t\hat{A}\otimes\hat{\sigma}_{y}}|\Psi_{i}\rangle_{sp} \\
=&\alpha|0\rangle_{s}\otimes e^{-i\theta\hat{\sigma}_{y}}|0\rangle_{p} + \beta|1\rangle_{s}\otimes e^{i\theta\hat{\sigma}_{y}}|0\rangle_{p}  \\
=&\alpha|0\rangle_{s}(\mathrm{cos}\theta|0\rangle_{p}+\mathrm{sin}\theta|1\rangle_{p})+\beta|1\rangle_{s}(\mathrm{cos}\theta|0\rangle_{p}-\mathrm{sin}\theta|1\rangle_{p}),
\end{split}
\end{equation}
where $\hbar\equiv 1$ and $\theta=g\Delta t$. Different eigenstates of observable $\hat{A}$ cause different rotation of pointer in Bloch sphere after interaction. When $\theta=\pi/4$, the two pointer states become $|\pm\rangle_{p}=(|0\rangle_{p}\pm |1\rangle_{p})/\sqrt{2}$ and they are orthogonal. In this case, if measurement is performed on pointer with basis $\lbrace|+\rangle, |-\rangle\rbrace$ we can definitely know the state of system after measurement, which corresponds to projective measurement. In general cases that $\theta\neq \pi/4$, we just realize POVM measurement and there exists ambiguity about state of system after measurement on the pointer. 

Weak measurement, however, focuses on weak interaction case of $\theta\ll 1$ in which barely no information about system can be extracted by measurement of pointer. The essential difference between weak measurement and conventional measurement is the introduction of post-selection of system. Suppose that the system is post-selected into the state $|\psi_{f}\rangle=\gamma|0\rangle+\eta|1\rangle$ after system-pointer interaction. The state of pointer, after post-selection, becomes
\begin{equation}
|\varphi\rangle_{p}=\frac{1}{N} \langle\psi_{f}|\Psi_{f}\rangle_{sp}=\frac{1}{N}     \langle\psi_{f}|e^{-i\theta\hat{A}\otimes\hat{\sigma}_{y}}|\psi_{i}\rangle_{s}|0\rangle_{p},
\end{equation}
where $N$ is normalization factor that $N=|\langle\psi_{f}|\Psi_{f}\rangle_{sp}|$. Since $\theta\ll 1$ in weak measurement, to the first order of approximation, the state of pointer is
\begin{equation}
\begin{split}
|\varphi\rangle_{p}&=\frac{\langle\psi_{f}|\psi_{i}\rangle}{N}e^{-i\theta\langle\hat{A}\rangle_{w}\hat{\sigma}_{y}}|0\rangle_{p}.  \\
\end{split}
\end{equation}
Here so-called weak value of observable $\hat{A}$ is defined as $\langle\hat{A}\rangle_{w}=\langle\psi_{f}|\hat{A}|\psi_{i}\rangle/\langle\psi_{f}|\psi_{i}\rangle$. Intuitively, post-selection of the system causes about $\theta\langle\hat{A}\rangle_{w}$ rotation of pointer on Bloch sphere. When proper measurement is performed on pointer, information about $\theta$ or weak value $<\hat{A}>_{w}$ is obtained. Specifically, we have 
\begin{equation}
\langle\hat{\sigma}_{+}\rangle_{p}=2\theta \mathrm{Re}\langle\hat{A}\rangle_{w},  \langle\hat{\sigma}_{R}\rangle_{p}=2\theta \mathrm{Im}\langle\hat{A}\rangle_{w},
\end{equation}
where $\hat{\sigma}_{+}\equiv|+\rangle\langle+|-|-\rangle\langle -|, \hat{\sigma}_{R}\equiv|R\rangle\langle R|-|L\rangle\langle L|$ with $|\pm\rangle = (|0\rangle \pm |1\rangle)/\sqrt{2}, |R/L\rangle=(|0\rangle \pm i|1\rangle)/\sqrt{2}$.


When $\langle\psi_{f}|\psi_{i}\rangle$ approaches to zero, $\langle\hat{A}\rangle_{w}$ becomes extremely large, signal $\theta$ is thus amplified. This is, of course, at the cost of extremely low success probability of $|\langle\psi_{f}|\psi_{i}\rangle|^{2}$. Weak value amplification (WVA) has been extensively studied since the observation of the spin hall effect of light \cite{Hall}. Due to the low successful probability of post-selection in the WVA, controversy on whether WVA measurement outperforms conventional measurement in parameter estimation has been last for decades. In most precision measurement experiments e.g., gravitational wave detection we actually care only about sensitivity, which cannot be simply improved by amplification.   
The fact that WVA has the potential to magnify ultra-small signals to the measurable level undoubtedly makes it useful. This is because the measurement apparatus in practice are limited to finite precision. The most vivid example is microscope, which is used to distinguish two spots that are invisible to human eyes. Microscope amplifies the spots so that we can direct see them for distinguish. However, the resolution of microscope is determined by scattering limit of light. Amplification helps us to distinguish objects with eyes but will not improve the resolution. More works related to weak measurements and its application can be found in references \cite{WV2, WV3, WV4, WV5,WV6, WV7} and references there.

Based on above discussion, application of WVA to gravitational wave detection will not improve the sensitivity of detector. Besides, how to amplify general longitudinal phase signal as in gravitational wave detection via WVA has not been resolved completely. We will show here that both of concerns can be eliminated. It should be realized that the definition of weak value is due to the choice of von Neumann-type interaction. Without this restriction we consider the control-rotation evolution 
\begin{equation}
\hat{U} = |0\rangle_{s}\langle 0|\otimes\hat{I}_{p} + |1\rangle_{s}\langle 1|\otimes(|0\rangle_{p}\langle 0|+e^{i\theta}|1\rangle_{p}\langle 1|)
\end{equation}
with $\theta$ be the phase signal to be measured. Choosing the initial state of pointer as $|+\rangle_{p}$ then the state of the composite system, after interaction, becomes
\begin{equation}
|\Psi_{f}\rangle_{sp}=\alpha|0\rangle_{s}\otimes|+\rangle_{p}+\beta|1\rangle_{s}\otimes(|0\rangle_{p}+e^{i\theta}|1\rangle_{p})/\sqrt{2}.
\end{equation}
When the system is in state of $|0\rangle_{s}$ nothing happens to pointer, while in state of $|1\rangle_{s}$ the state of pointer is rotated $\theta$ along the equator of Bloch sphere. With post-selection of the system into state $|\psi_{f}\rangle$, the pointer state reads (unnormalized)
\begin{equation}
|\varphi\rangle_{p}=\langle\psi_{f}|\Psi_{f}\rangle_{sp}=(\alpha\gamma+\beta\eta)|0\rangle_{p}+(\alpha\gamma+\beta\eta e^{i\theta})|1\rangle_{p}
\end{equation}
with $\alpha, \beta, \gamma, \eta$ are taken real numbers here. $\alpha\gamma+\beta\eta e^{i\theta}$ can be recast as $\sqrt{\alpha^{2}\gamma^{2}+\beta^{2}\eta^{2}+2\alpha\gamma\beta\eta\mathrm{cos}\theta}e^{i\phi}$ with
\begin{equation}
\mathrm{tan}\phi=\frac{\beta\eta\mathrm{sin}\theta}{\beta\eta\mathrm{cos}\theta+\alpha\gamma}.
\end{equation}
Since phase signal $\theta\ll 1$, $\sqrt{\alpha^{2}\gamma^{2}+\beta^{2}\eta^{2}+2\alpha\gamma\beta\eta\mathrm{cos}\theta}=\alpha\gamma+\beta\eta$ in the first order approximation and the state of pointer thus becomes
\begin{equation}
|\varphi\rangle_{p}=\frac{1}{\sqrt{2}}(|0\rangle_{p}+e^{i\phi}|1\rangle_{p}).
\end{equation}
Post-selection of the system after weak interaction results in $\phi$ rotation of the pointer state along equator of Bloch sphere. Amplification is realized when we choosing post-selection state $|\psi_{f}\rangle$ properly such that $\phi > \theta$ is satisfied. In general, we can set $\alpha=\beta=1/\sqrt{2}$ and $\gamma=\mathrm{cos}\chi, \eta=\mathrm{sin}\chi$, in the case of $\theta\ll 1$, Eq. (21) reduces to
\begin{equation}
\mathrm{tan}\phi =\frac{\theta}{1+\mathrm{cot}\chi}.
\end{equation}
Suppose that $\chi=-(\pi/4+\delta)$ with $\delta\ll 1$, then $\mathrm{cot}\chi=-1+\delta$ in first order approximation and we have $\mathrm{tan}\phi=\theta/\delta$.
The factor of amplification in this case is
\begin{equation}
h=\frac{\phi}{\theta}=\frac{\mathrm{arctan}(\theta/\delta)}{\theta}.
\end{equation}
The amplified phase $\phi$ can be easily extracted by performing $\hat{\sigma}_{R}$ measurement on pointer
\begin{equation}
\langle\hat{\sigma}_{R}\rangle= _{p}\langle\varphi|\hat{\sigma}_{R}|\varphi\rangle_{p}=\mathrm{sin}\phi.
\end{equation}


The above amplification protocol is within weak measurement framework but different from WVA and thus named as weak measurement amplification (WMA). The concept of weak value is not required in WMA.
In analogy with micrometer that transforms a small displacement into a larger rotation of circle, WMA transforms a ultra-small phase signal into a larger rotation of the pointer state along equator of Bloch sphere. 
The WMA is a general amplification protocol and applied to any physical qubit systems. 

We have figured out how to realize longitudinal phase amplification within weak measurement framework. The next natural question is how to apply WMA to gravitational wave detection or more specifically how to construct detection system based on WMA protocol. Fortunately, it is not hard to construct an optical system realize WMA. As shown in Fig. 3, the path and polarization degrees of freedom of photons are chosen as system and pointer respectively and polarization Michelson interferometer is used to realize control-rotation evolution. 

\begin{figure}[tbp]
\centering
\includegraphics[scale=0.40]{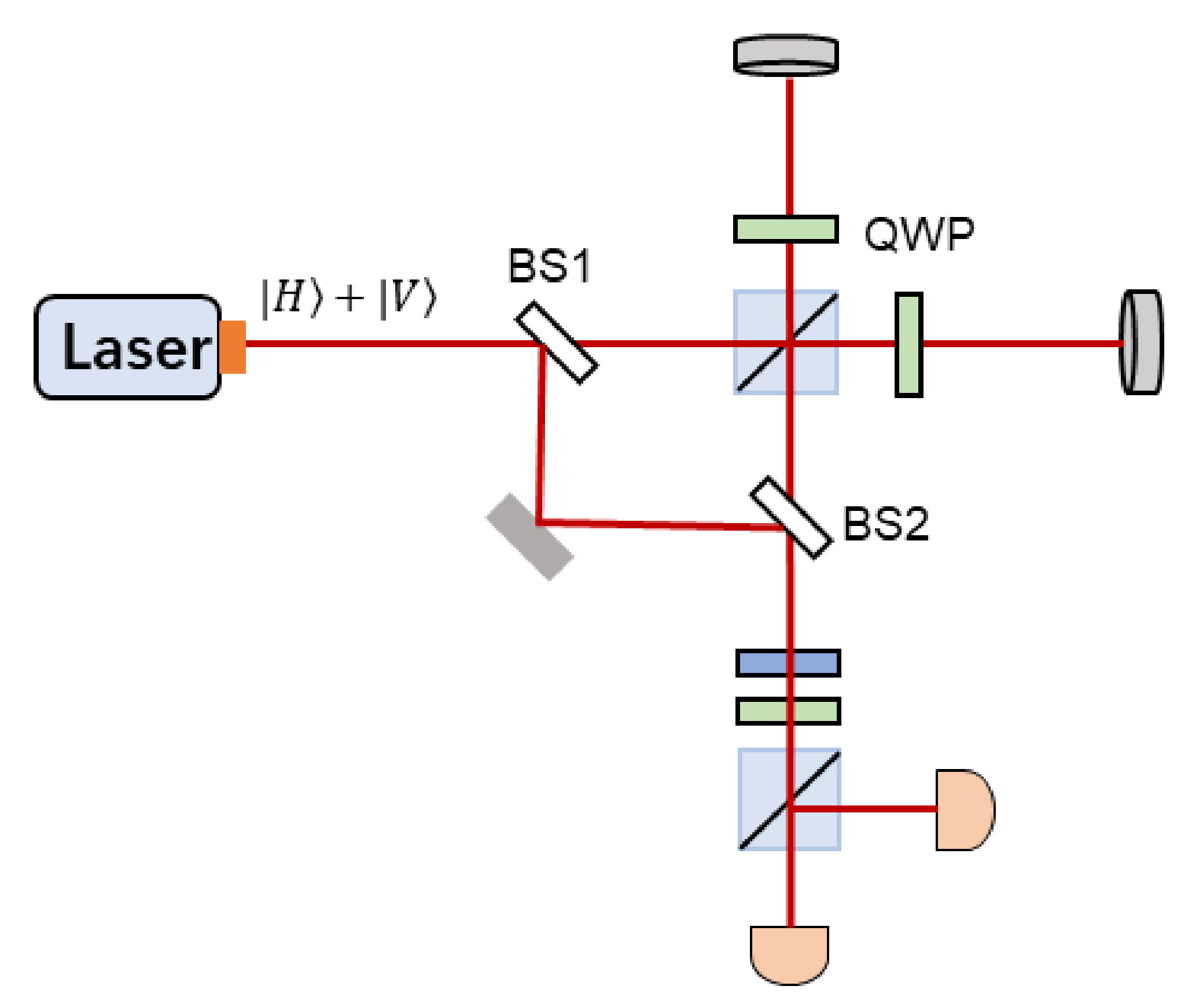}
\caption{ Optical realization of weak measurement amplification. This optical configuration is the foundation of proposed detector in main text.}
\label{f4}
\end{figure} 

The WMA optical system consists of five parts i.e., laser source, initial state preparation, control-rotation evolution, post-selection and pointer measurement. The laser source produces stabilized photons in the linear polarization state $|+\rangle=(|H\rangle+|V\rangle)/\sqrt{2}$. The initial state preparation is
fulfilled by the beam splitter (BS1). The state of photons, after passing through the BS1, becomes
\begin{equation}
|\Psi_{i}\rangle_{sp}=|\psi_{i}\rangle_{s}\otimes|+\rangle_{p}=(r_{1}|d\rangle+t_{1}|u\rangle)\otimes|+\rangle_{p},
\end{equation}
where $r_{1}, t_{1}$ are coefficients of reflection and transmission of the BS1 satisfying $|r_{1}|^{2}+|t_{1}|^{2}=1$ and $|u\rangle, |d\rangle$ represent path state of up arm and down arm respectively. The photons, which fly along path of up arm, enters a polarization Michelson interferometer (PMI) that used for signal collection. The PMI, which consists of a polarizing beam splitter (PBS1), two quarter wave plates (QWP) and two end masses, outputs the state of photons $(|H\rangle+e^{i\theta}|V\rangle)/\sqrt{2}$ when the input state is $(|H\rangle+|V\rangle)/\sqrt{2}$ with $\theta$ is the phase signal to be measured. The function of QWP, which is fixed at $\pi/4$, is to transform polarization states $|H\rangle$ and $|V\rangle$ into its orthogonal states $|V\rangle$ and $|H\rangle$ respectively when photons
pass through it twice. Hence the transmitted photons with the state $|H\rangle$ is converted by QWP to the state $|V\rangle$, which is reflected by the PBS1, and the reflected
photons with the state $|V\rangle$ is converted to the state $|H\rangle$ by the other QWP and is thus transmitted by the PBS1 such that the photons will not come out from the input port. There is no change of state when photons flying along the down arm path. The control-rotation interaction is fulfilled when photons come out of PMI with state
\begin{equation}
|\Psi_{f}\rangle_{sp}=r_{1}|d\rangle\otimes|+\rangle+t_{1}|u\rangle\otimes(|H\rangle+e^{i\theta}|V\rangle)/\sqrt{2}.
\end{equation}
The process of signal amplification, which depends on post-selection, can be fulfilled by the BS2 with coefficients of reflection and transmission of $r_{2}$ and $t_{2}$, which
satisfy $|r_{2}|^{2}+|t_{2}|^{2}=1$. The post-selection is completed when we focus only on the photons come out from the down port of BS2 with the post-selected path state
\begin{equation}
|\psi_{f}\rangle_{s}=r_{2}|d\rangle+t_{2}|u\rangle.
\end{equation}
The polarization state of post-selected photons, in the first order approximation, thus becomes
\begin{equation}
|\varphi\rangle_{p}=\frac{1}{\sqrt{2}}(|H\rangle+e^{i\phi}|V\rangle)
\end{equation}
with $\phi$ determined by
\begin{equation}
\mathrm{tan}\phi=\frac{\mathrm{sin}\theta}{\mathrm{cos}\theta+r_{1}r_{2}/t_{1}t_{2}}.
\end{equation}
The amplified phase signal $\phi$ thus can be obtained by properly choosing $r_{1}, r_{2}, t_{1}, t_{2}$ such that $_{s}\langle\psi_{f}|\psi_{i}\rangle_{s}=r_{1}r_{2}+t_{1}t_{2}\to 0$. The HWP, QWP, PBS2 and two detectors complete the detection of amplified phase signal as a polarization analyser.

\section{Appendix B: Squeezing Quantum Noise}
\begin{figure}[tbp]
\centering
\includegraphics[scale=0.40]{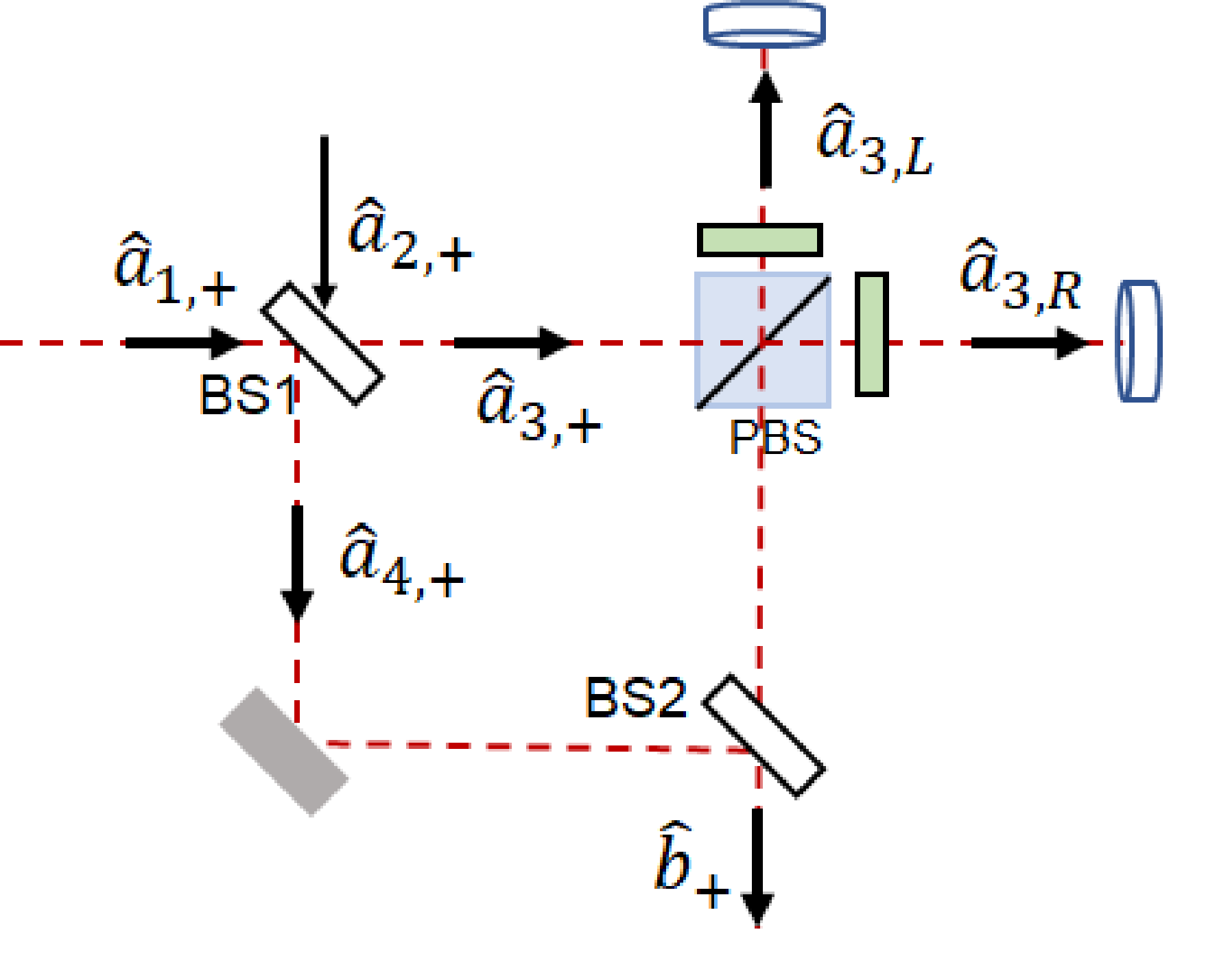}
\caption{input-output of quantized optical field.  }
\label{f3}
\end{figure}

In this section, we will show that the quantum noise in proposed detector results from the vacuum fluctuations in the dark port of BS1. The analysis follows the way of Caves \cite{Caves}. The quantized light field propagating in the direction of {\it z}  with $|+\rangle=(|H\rangle+|V\rangle)/\sqrt{2}$ polarization can be expressed as
\begin{equation}
\hat{E}_{+}=\sqrt{\frac{2\pi\hbar \omega}{Ac}}[\hat{a}_{+}e^{-i(\omega t-kz)}+\hat{a}_{+}^{\dagger}e^{i(\omega t-kz)}]
\end{equation}
where $A$ is cross-section of light field and $\hat{a}_{+}, \hat{a}^{\dagger}_{+}$ represent annihilation and creation operator of photons with $|+\rangle$ polarization. Clearly, we have $\hat{a}_{+}=(\hat{a}_{H} + \hat{a}_{V})/\sqrt{2}$.
First, consider the dark port output of BS2 as shown in Fig. \ref{f3}. For BS1, the input-output relations are
\begin{equation}
\begin{split}
\hat{a}_{3,+}&=t_{1}\hat{a}_{1,+} + r_{1}\hat{a}_{2,+} \\
\hat{a}_{4,+}&=t_{1}\hat{a}_{2,+} + r_{1}\hat{a}_{1,+}.
\end{split}
\end{equation}
Here it is assumed that all optical elements are ideal and optical losses can be neglected.  For simplicity of discussion, we assume that there exists no signals and other noises such that the polarization state of photons is unchanged after passing through the polarization Michelson interferometer. The dark port output of BS2 is thus
\begin{equation}
\begin{split}
\hat{b}_{+}&=t_{2}\hat{a}_{3,+} + r_{2}\hat{a}_{4,+} \\
&=(t_{1}t_{2}+r_{1}r_{2})\hat{a}_{1,+} + (t_{1}r_{2}+r_{1}t_{2})\hat{a}_{2, +}
\end{split}
\end{equation}
Suppose that there are certain $N$ photons are input from the port 1 of BS1, then the initial state of the whole system is $|N\rangle_{1,+}\otimes|0\rangle_{2,+}$. The average number of photons coming out of dark port of BS2 is given as
\begin{equation}
\langle\hat{n}\rangle=\langle\hat{b}^{\dagger}\hat{b}\rangle=|t_{1}t_{2}+r_{1}r_{2}|^{2}N=PN.
\end{equation}
It can be seen that $P$ is the success probability of post-selection. The direct calculation gives the fluctuation of photons number as
\begin{equation}
\begin{split}
\Delta n &=\sqrt{\langle\hat{n}^{2}\rangle-\langle\hat{n}\rangle^{2} } \\
&=\sqrt{|t_{1}t_{2}+r_{1}r_{2}|^{2}|t_{1}r_{2}+r_{1}t_{2}|^{2}}\cdot\sqrt{N}.
\end{split}
\end{equation}
Note that only terms like $\hat{a}_{1}^{\dagger}\hat{a}_{1}\hat{a}_{1}^{\dagger}\hat{a}_{1}$ and $\hat{a}_{1}^{\dagger}\hat{a}_{1}\hat{a}_{2}\hat{a}_{2}^{\dagger}$ make contributions.
In our case, $t_{1}=1/\sqrt{2}, r_{1}=i/\sqrt{2}$ and $t_{2}\approx 1/\sqrt{2}, r_{2}\approx i/\sqrt{2}$ with $\delta\ll 1$, then $|t_{1}r_{2}+r_{1}t_{2}|^{2}\approx 1$, we obtain $\delta n=\sqrt{PN}$ as expected. 
The shot noise is indeed due to the vacuum fluctuation of dark port of BS1.

\begin{figure*}[tbp]
\centering
  \subfigure[]{
    \includegraphics[scale=0.22]{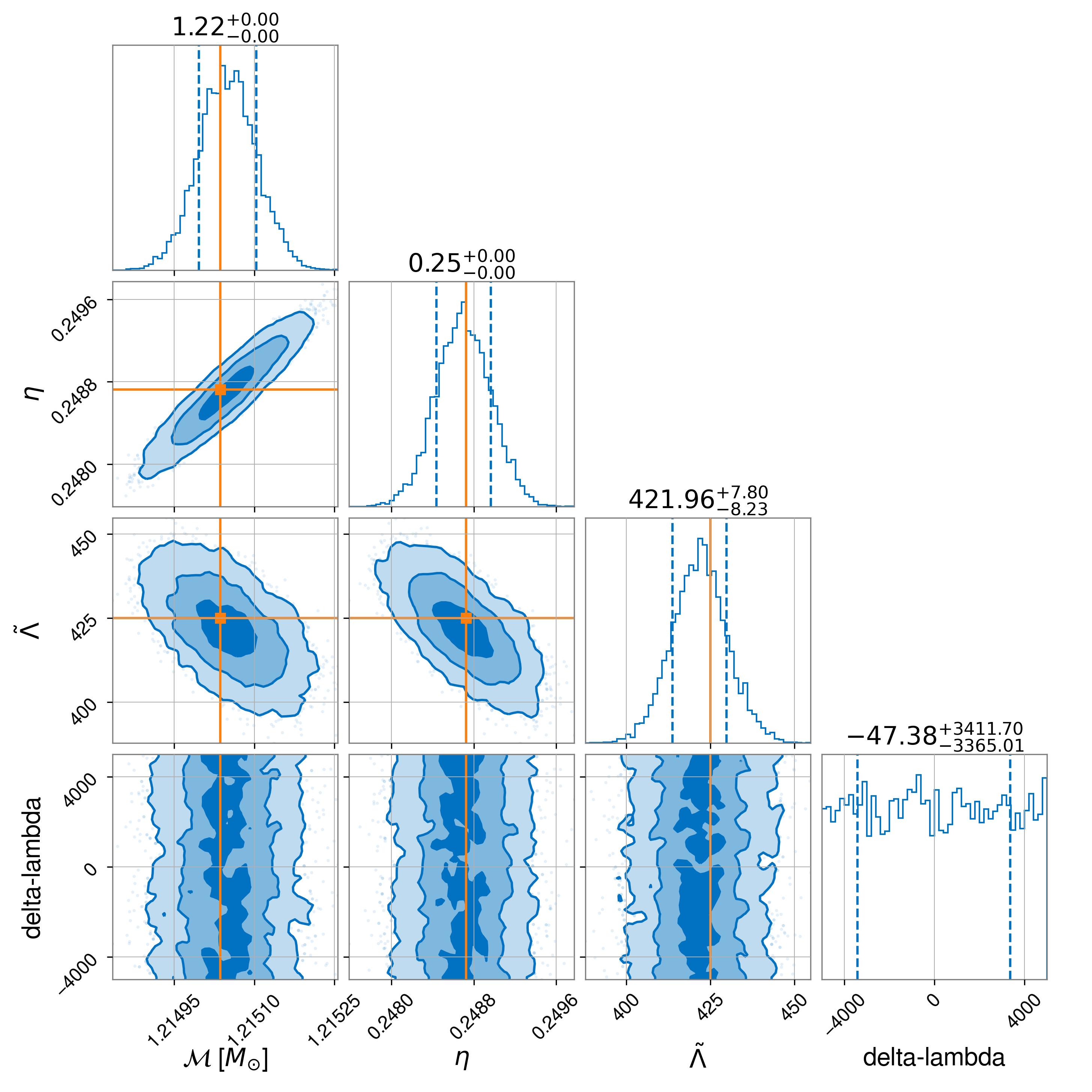}}
  \hspace{0.05in} 
  \subfigure[]{
    \includegraphics[scale=0.22]{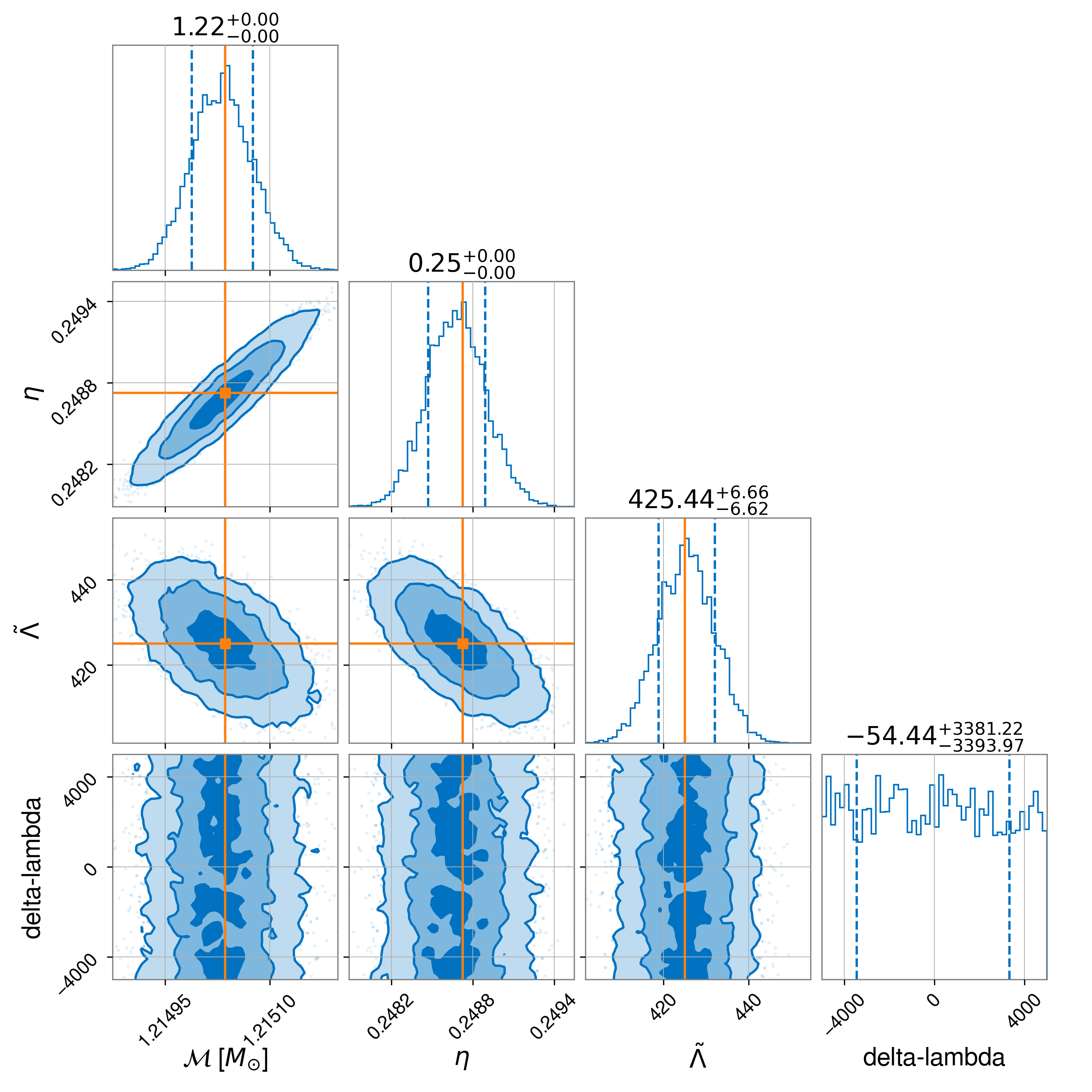}}
  \hspace{0.05in}
  \subfigure[]{
    \includegraphics[scale=0.22]{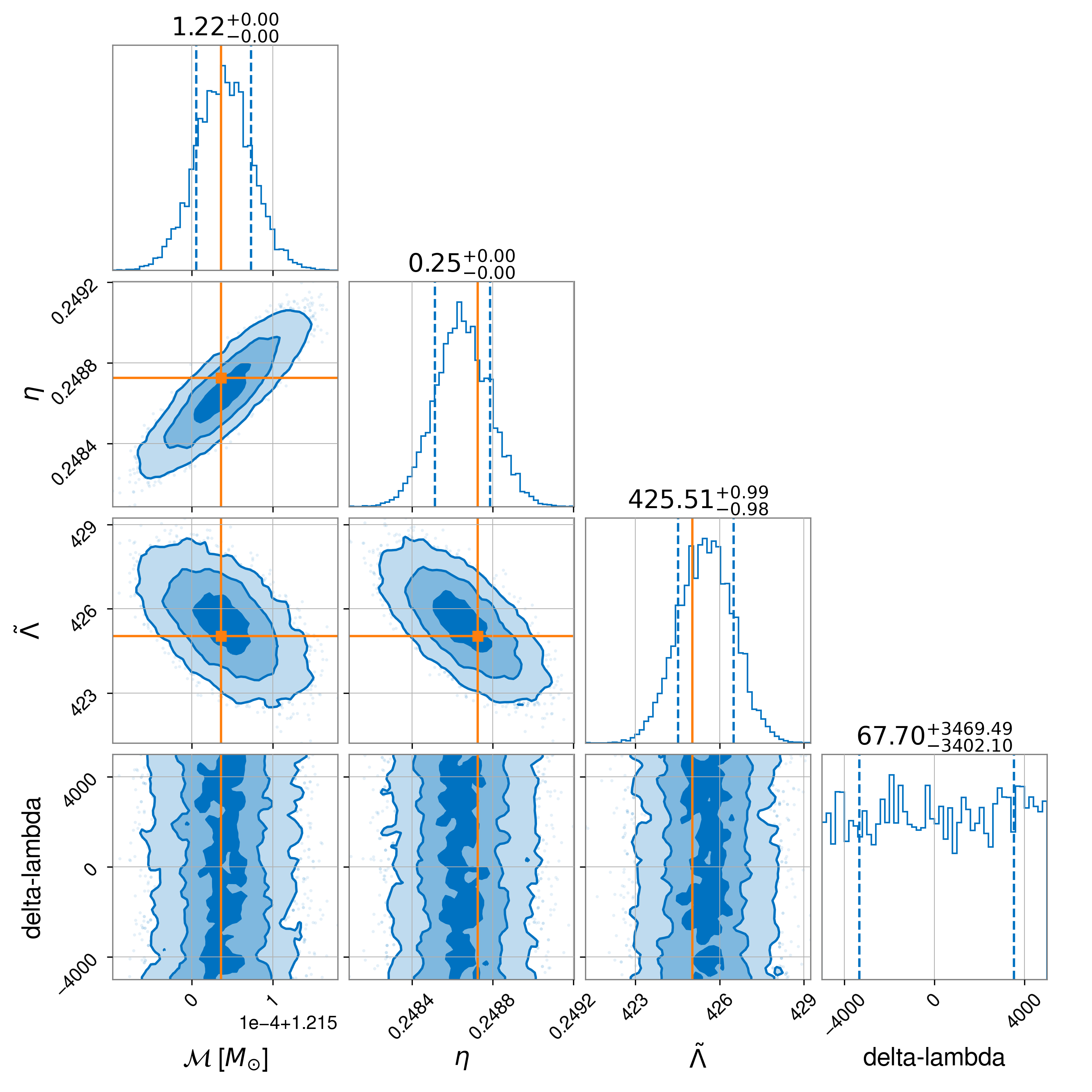}}
  \caption{Comparison of parameters estimation with (a) Network A; (b) Network B; and (c) Network C. }
\label{f5}
\end{figure*} 

\begin{figure*}[tbp]
\centering
  \subfigure[]{
    \includegraphics[scale=0.18]{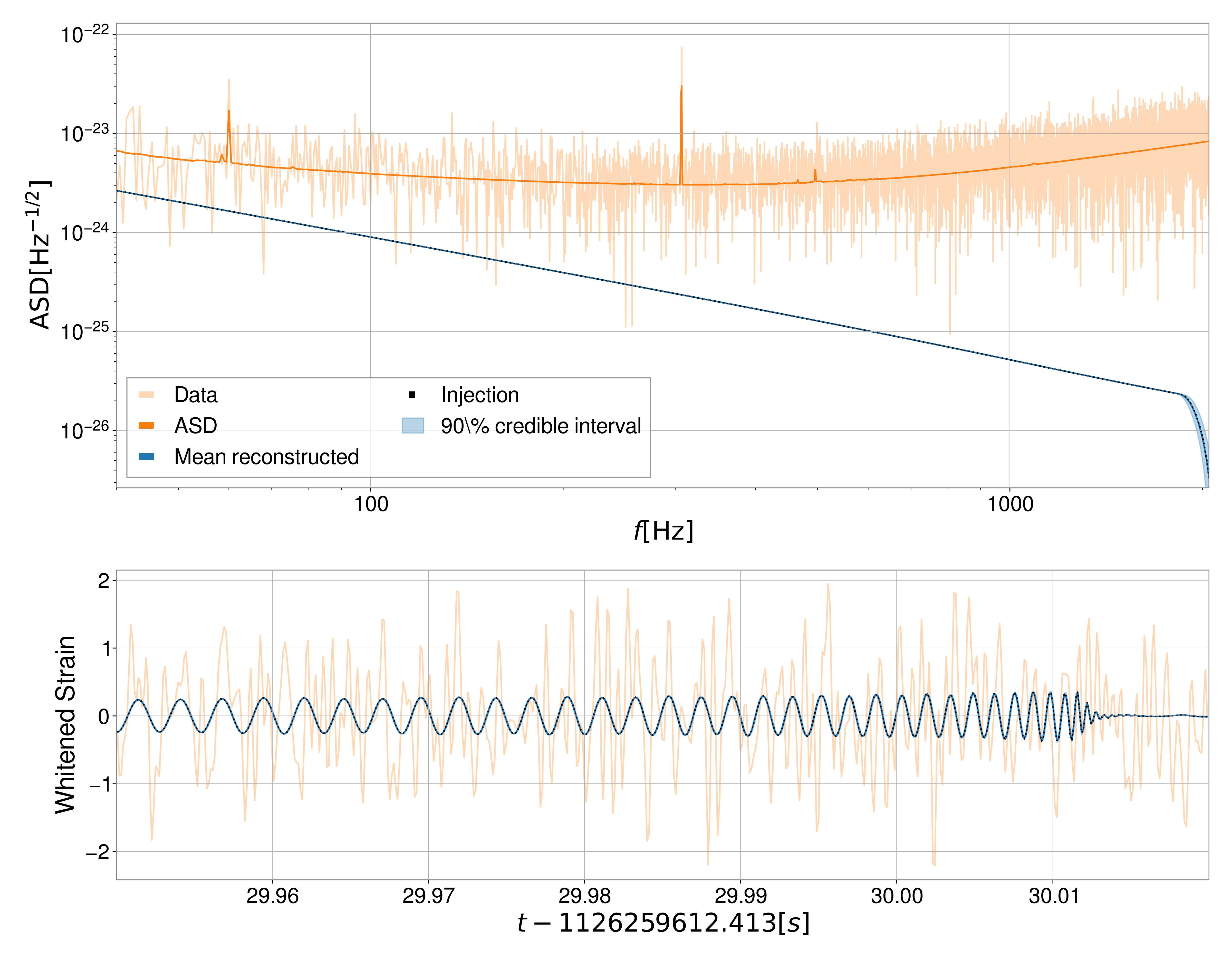}}
  \hspace{0.05in} 
  \subfigure[]{
    \includegraphics[scale=0.18]{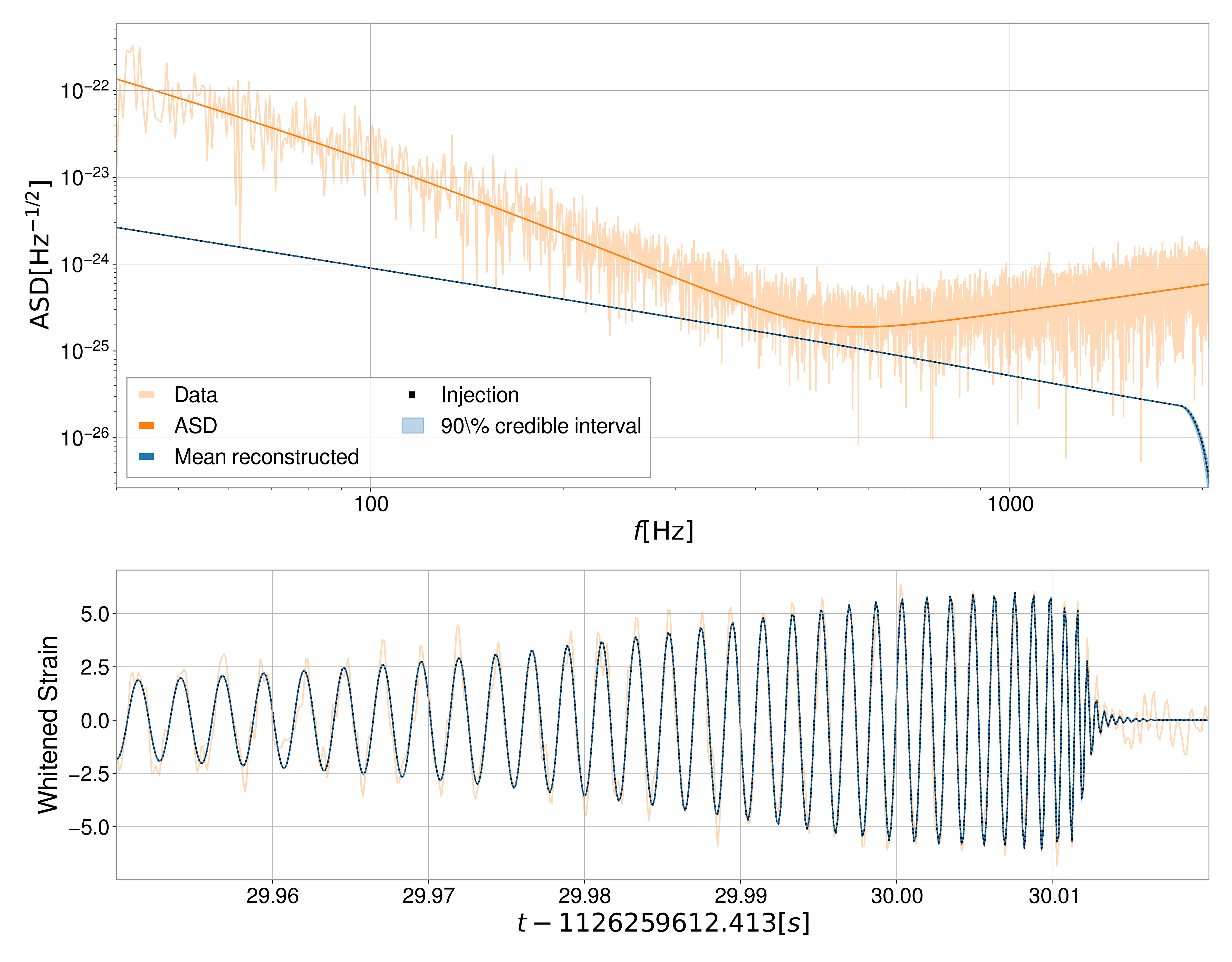}}
  \caption{Waveform reconstruction of (a) LIGO-like detector in Network B; and (b) Proposed detector in Network C. }
\label{f6}
\end{figure*} 

We now consider the radiation pressure noise that originates from photons impinging on mirrors. The optical field in the two arms $\hat{a}_{3, R}, \hat{a}_{3, L}$ is related to $\hat{a}_{3}$ as
\begin{equation}
\hat{a}_{3,+} =\frac{1}{\sqrt{2}}(\hat{a}_{3,H}+\hat{a}_{3,V})= \frac{1}{\sqrt{2}}(\hat{a}_{3,R}+\hat{a}_{3,L}).
\end{equation}
According to Eq. (32), we have
\begin{equation}
\begin{split}
\hat{a}_{3,R} &= t_{1}\hat{a}_{1,H} + r_{1}\hat{a}_{2,H} \\
\hat{a}_{3,L} &= t_{1}\hat{a}_{1,V} + r_{1}\hat{a}_{2,V}.
\end{split}
\end{equation}
What we care about is difference of photons number in two arms $\delta\hat{n}=\hat{a}_{3,R}^{\dagger}\hat{a}_{3,R}-\hat{a}_{3,L}^{\dagger}\hat{a}_{3,L}$. 
Notice that $\langle\hat{a}_{1,H}^{\dagger}\hat{a}_{1,H}\rangle=\langle\hat{a}_{1,V}^{\dagger}\hat{a}_{1,V}\rangle=N/2$. The average value of number difference is $\langle\delta\hat{n}\rangle=0$.
The fluctuation $\Delta(\delta\hat{n})$ of number difference is thus determined by
\begin{equation}
\Delta(\delta\hat{n}) =\sqrt{\langle(\delta\hat{n})^{2}\rangle} =t_{1}\sqrt{N}.
\end{equation}

We have shown that quantum noise in the proposed detector results from the vacuum fluctuations of dark port of BS1 and thus usage of squeezed light in dark port of BS1 will certainly reduce shot noise or radiation pressure noise.

\section{Appdendix C: Comparison of Detection Networks }

In this section, we provide more details of comparison of different detection networks in one of which the proposed detector is included. We assume that the proposed detector with $4\mathrm{km}$ arm length is deployed in the Hefei, Anhui Province of China, with location information as latitude 117.25, longitude 31.83 and elevation 51.884. The azimuth of two arms are chosen to be X-arm 70.5674 and Y-arm 160.5674. 
The detection networks to be compared includes A: Current network with two LIGO detectors and Virgo detector, B: Current network plus LIGO-like detector located in Hefei and C: Current network plus proposed detector located in Hefei. 

With open source software Bilby, we simulate a $1.5M_{\bigodot}-1.3M_{\bigodot}$ BNS located at a distance of $100\mathrm{Mpc}$ with IMRPhenomD$\_$NRTidal waveform is used. 
For BNS in pre-merger stage, emitted gravitational wave contains the tidal deformation information, which can be used to study the equation of state of neutron star. The tidal deformability of neutron star with mass $m$ is defined as
\begin{equation}
\Lambda = \frac{2k_{2}}{3}(\frac{c^{2}R}{Gm})^{5},
\end{equation}
where $R$ is the radius, and $k_{2}$ is the second Love number that measures the rigidity of the neutron star. Both $R$ and $k_{2}$ are determined by the equation of state of neutron star with fixed mass. In our simulation we set $\Lambda_{1}=400$ for $1.5M_{\bigodot}$ neutron star and $\Lambda_{2}=450$ for $1.3M_{\bigodot}$ neutron star. Since $\Lambda_{1}$ and $\Lambda_{2}$ are highly correlated, it is more convenient to consider dimensionless tidal deformability $\widetilde{\Lambda}$ and $\delta\widetilde{\Lambda}$ defined in Ref. \cite{deform}. The $\lbrace\Lambda_{1}, \Lambda_{2}\rbrace$ can be parametrized in terms of $\lbrace\widetilde{\Lambda}, \delta\widetilde{\Lambda}\rbrace$. 
The inject waveform signal is determined by total $15$ relevant parameters including masses, distance and tidal deformablities. The rest parameters in our simulation is chosen as the same as the $binary_{-}neutron_{-}star_{-}example$ given in Bilby software package. Bilby uses Bayesian inference method to estimate relevant parameters. In order to reduce inference time and focus on parameters interesting for BNS merger physics, only chirp mass $\mathcal{M}$, symmetric mass ratio $\eta$, and $\widetilde{\Lambda}, \delta\widetilde{\Lambda}$ are sampled. We use a Gaussian prior distribution in chirp mass between $0.1 < \mathcal{M} < 1.215$ and use uniform prior distribution for other three parameters with $0.1 < \eta < 0.25, 0 < \widetilde{\Lambda} < 5000, -5000 < \delta\widetilde{\Lambda} < 5000$. 
  
Fig. \ref{f5} shows the corner plot for three different detection networks. All detection networks are not able to discern the $\delta\widetilde{\Lambda}$ contribution. This is expected due to $\delta\widetilde{\Lambda}$ only shows up in 6 post-Newtonian tidal correction. The injected value of $\mathcal{M}, \eta$ are well recovered. For the tidal deformability $\widetilde{\Lambda}$, network B and C recover better than network A. This is because the combined SNR $\rho$ of network B and C are improved by adding extra detector in network A.  The SNR of individual detector is defined as 
\begin{equation}
\rho_{i}^{2}=4\int\frac{|\widetilde{h}(f)|^{2}}{S_{h}(f)}df,
\end{equation}
and the combined SNR of detection network is given by $\rho=\sqrt{\sum_{i}\rho_{i}^{2}}$. Since tidal effect becomes important in the late-inspiral phase of BNS at frequencies around $1k\mathrm{Hz}$, Network C including the proposed detector behaves much better than network A and B in the inferred uncertainty of tidal deformablity $\Delta\widetilde{\Lambda}$. The fractional error in the tidal deformability defined by $\Delta\widetilde{\Lambda}/\widetilde{\Lambda}$ quantifies how good we constraint equation of state, and it is shown in TABLE I that network C preforms approximately six times better than other two networks.
Fig. \ref{f6} shows the reconstructed signals for the LIGO-like detector in network B and the proposed detector in network C respectively. The yellow curves simulate the signals(injected signal plus noise signal) recorded by detectors and the blue curves show the reconstructed gravitational wave signals with $90\%$ credible interval. Since the network C has higher combined SNR, it can be seen from (b) of Fig. \ref{f6} that gravitational wave signal can be more easily be identified and reconstructed. After the merger of BNS, a neutron star may survives and emits gravitational waves at frequencies of $1\mathrm{k}-4\mathrm{kHz}$ for up to hundreds of milliseconds. The proposed detector with sensitivity enough in this frequency domain can help us reconstruct the post-merger signal more accurately and thus provides us opportunity to probe the equation of state of neutron star in a different density regime \cite{detector6}. The comparison of (a) and (b) in Fig. \ref{f6} show clearly the advantage of adding the proposed detector in current detection network.

\end{document}